\documentclass[pra,twocolumn,showpacs,aps,floatfix,amsmath,amssymb,amsfonts]{revtex4-1}

\usepackage{amsmath}
\usepackage{graphicx}
\usepackage{color}

\newcommand{\beq}{\begin{equation}}
\newcommand{\eeq}{\end{equation}}

\begin{document}

\title{Precise Feshbach resonance spectroscopy using tight anharmonic traps}
\author{Krzysztof Jachymski}
\affiliation{
Institute for Quantum Control (PGI-8), Forschungszentrum J\"ulich, D-52425 J\"{u}lich, Germany}
\date{\today}

\begin{abstract}
Feshbach resonances are among the essential control tools used in ultracold atom experiments. However, for complex atomic species the theoretical characterization of resonances becomes challenging. For closely spaced resonances, the measurement of three-body losses does not provide sufficient resolution to discriminate them. For this reason, resonance spectroscopy of trapped isolated atoms is becoming the state of the art. Here we show that trapping the atoms in a double well potential such as an optical lattice or a pair of optical tweezers enables precise characterization of not only the resonance position and width, but also its pole strength, giving valuable information about the atomic structure relevant for subsequent many-body studies.

\end{abstract}

\maketitle
\section{Introduction}
 
Ultracold trapped atoms provide clean quantum systems with remarkable control possibilities that can be used for quantum technological applications. Much of the success of these systems can be attributed to the existence of Feshbach resonances (FR), which allow for tuning the interaction strength between the atoms~\cite{JulienneRMP}. Identification of Feshbach resonances is thus an essential early step in experiments, especially with new atomic species~\cite{Julienne2009,Takekoshi2012,Berninger2013}. Recent works~\cite{Mark2018,hood2019multichannel} have demonstrated that confining the atoms in an external trap increases the experimental precision of Feshbach spectroscopy significantly by getting rid of three-body effects such as the ones related to Efimov physics. Moreover, the resonances enable the association of weakly bound molecular states which is often the crucial step in creation of ultracold molecules~\cite{Kohler2006}. These are potentially very promising for a multitude of future applications~\cite{Carr2009,PSJ2012}. As direct cooling of molecules to quantum degeneracy is still in the early stages of development~\cite{Prehn2016,Norrgard2016}, association of ultracold atoms into an already cold molecular product can be more efficient. The schemes that result in systems with minimal temperature and entropy rely on trapping the atoms in an optical lattice in a double Mott insulator state with two atoms in each lattice site~\cite{Moses2015,Covey2016}. Recently developed systems in which single atoms are kept in optical tweezers~\cite{Liu2018} allow for realization of similar controlled collision protocols at a single atom level. 

While for pairs of alkali atoms it is usually possible to contruct a very precise theoretical model which fully captures the resonance spectrum with only a handful of free parameters to fit, new systems have recently emerged that cannot be treated that way. Collisions of lanthanide atoms with high magnetic moments such as erbium and dysprosium involve hundreds of coupled channels due to large interaction anisotropy, leading to multiple resonances per gauss~\cite{Maier2015,Baier2018}. Similar situation occurs for atom-molecule and molecule-molecule collisions~\cite{Mayle2012}, where resonances have recently been observed~\cite{Yang2019}. Coupled multichannel systems require extensive numerical resources and cannot provide classification of bound states in terms of quantum numbers. For this reason, theoretical effort has so far been focused on simplified effective models as well as studying generic scattering properties of strongly coupled multichannel systems~\cite{Mayle2012,Jachymski2015,Croft2017,Makrides2018,Mehta2018}.

Particles with rich internal structure can also be trapped in a lattice to improve the efficiency of their quantum state manipulation. The interactions between ultracold atoms are usually described using the scattering length as a single parameter. This is sufficient as long as the length and energy scales associated with the trap are much larger than the characteristic interaction energy and the gas is dilute and cold enough that the $s$-wave scattering dominates completely. However, the energy dependence of the scattering phase shift, which can be described e.g. in terms of the effective range, can become important even for ultracold systems. Single channel atomic interactions in harmonic traps are theoretically well understood~\cite{Bolda2002,Blume2002,Idziaszek2005}. This is important, as a single lattice site or tweezer can usually be approximated by the harmonic oscillator potential. Anharmonic corrections to the trapping potential couple the center of mass and relative motion and can lead to inelastic processes which have been observed in experiment~\cite{Sala2012,Sala2013}. Finite energy corrections can shift the resonance positions in the presence of the trap compared to the free space case~\cite{Naidon2007,Jachymski2017}. Furthermore, three-body recombination processes relevant for e.g. Efimov physics crucially depend not only on the scattering length, but also on the pole strength of the resonance~\cite{Ohara2012,Shotan2014,Wang2014}. From the many-body standpoint, momentum-dependent interactions around a narrow resonance can significantly alter  the physics of the system such as the Fermi gas across the BEC-BCS crossover~\cite{Gurarie2007}. Finally, multichannel interactions in optical lattice can give rise to novel effective lattice models~\cite{Doccaj2015,Wall2017}. Optical methods of studying the energy dependence of the resonance position have been recently proposed~\cite{Arunkumar2018}.

In this work, we consider FR in a double well trap and study the energy levels using a multichannel model developed in~\cite{Jachymski2013a}, which is capable of describing resonances with arbitrary pole strength. We focus on inelastic resonances induced by the anharmonic terms in the trapping potential, showing that differences between open- and closed-channel dominated FRs can arise. This shows that using double well traps for Feshbach resonance spectroscopy provides the tools to not only detect the positions of narrow resonances, but to fully characterize them even without an underlying theoretical model.

This work is structured as follows. In Section~\ref{sec2} we briefly describe the calculation method. The model trapping potential and its relation to optical lattices is the subject of Sec.~\ref{sectrap}. The resulting energy spectra are discussed in Sec~\ref{secres}. Conclusions are drawn in Sec.~\ref{secconc}.

\section{Formalism}
\label{sec2}
The calculations performed in this work are based on a minimal two-channel model with short-range interchannel coupling presented in more detail in~\cite{Jachymski2013a,Jachymski2016}. This approach partly relies on the separation of length scales between the interaction and the external trap~\cite{Bolda2002}, so that one can define the short-range boundary condition for the wave function independently of the confinement.
We consider the following Hamiltonian describing two trapped interacting ultracold atoms
\beq
\begin{split}
	H=\left|o\right>\left<o\right|\left(\frac{p^2}{2\mu}+\frac{P^2}{2M}+V(\mathbf{r},\mathbf{R})\right)+\\
	+\left|\chi\right>\left<\chi\right|\left(\frac{P^2}{2M}+\tilde{V}(\mathbf{R})\right)
	+W\left|\chi\right>\left<o\right|+\mathrm{h.c.}\, ,
\end{split}
\label{eq:ham}
\eeq

where $\left|o\right>$ and $\left|\chi\right>$ denote the open and closed (molecular) channel respectively, $\mu$ is the reduced mass of the pair, $M$ is the total mass, $V$ is the external trap, $p$ describes the relative momentum and $P$ stands for the center of mass momentum. Here the molecule is treated as a pointlike particle due to the length scale separation, so that the coupling $W(\mathbf{r})=g\delta(\mathbf{r})$ and the trapping potential in the closed channel $\tilde{V}(\mathbf{R})$ is obtained as $V(0,\mathbf{R})$. Note that the background interaction in the open channel has been neglected, as we focus on the effects of the resonance. We find the spectrum of the Hamiltonian by expanding the wave function in the basis of the single channel eigenstates. The presence of an external trap makes this approach convenient, as the required basis size is not large. Inserting the expansion of the form $\left|\Psi\right>=\left|o\right>\sum_{n}{c_{n}\psi_n(\mathbf{r},\mathbf{R})}+\left|\chi\right>\sum_{p}{a_p \Phi_p(\mathbf{R})}$ into the Hamiltonian~\eqref{eq:ham}, we obtain a self-consistent equation for the eigenenergies
\beq
\left(E-\epsilon_n\right)c_n=\sum_{jk}{\frac{\lambda_{kj}^\star \lambda_{kn}}{E-\nu-\varepsilon_k}c_j}
\eeq
with $\epsilon_n$ denoting the open channel eigenenergies, $\varepsilon_n$ the closed channel ones along with the shift $\nu$, and $\lambda_{kn}=g\langle\psi_n(\mathbf{r},\mathbf{R})|\delta(\mathbf{r})|\Phi_k(\mathbf{R})\rangle$ is the coupling matrix element.

However, in analogy with the free-space scenario the model has an ultraviolet divergence due to using the Dirac delta coupling without proper regularization and so renormalization of the parameters becomes necessary. Here we use the scheme developed for harmonic traps which is particularily simple and requires no numerical effort~\cite{DienerHo2006,Buchler,Jachymski2013a,Ilg2018}. In an anisotropic harmonic trap, it is sufficient to introduce the dressed resonance shift $\nu^\star=\nu-g^2\sqrt{\eta n^\star}/\pi$, where $\eta$ is the trap anisotropy and $n^\star$ is the number of considered basis states in the radial direction. The interchannel coupling $g$ and the shift $\nu^\star$ are linked to the actual resonance parameters via
\beq
 \nu^\star=\delta\mu\left(B-B_0\right)
 \eeq
 and
 \beq
  g=\hbar\omega\sqrt{2\pi a_{bg}\Delta\delta\mu/(d\hbar\omega)}\, .
  \label{gdef}
  \eeq
 Here the parameters describing the FR are its position $B_0$, width $\Delta$, background scattering length $a_{\rm bg}$ and the magnetic moment difference $\delta\mu$. One can then define a dimensionless quantity $s_{6}=a_{\rm bg}\Delta\delta\mu/(R_6 E_6)$ given in characteristic van der Waals units~\cite{vdW}, which describes the pole strength of the FR. The harmonic trap is described by its frequency $\omega$, anisotropy $\eta$ and characteristic width $d=\sqrt{\hbar/\mu\omega}$. This treatment can also be generalized to include multiple closed channels, either coupled or pre-diagonalized~\cite{Doccaj2015,Jachymski2016}.


\section{Double well trap}
\label{sectrap}

\begin{figure}
	\includegraphics[width=0.45\textwidth]{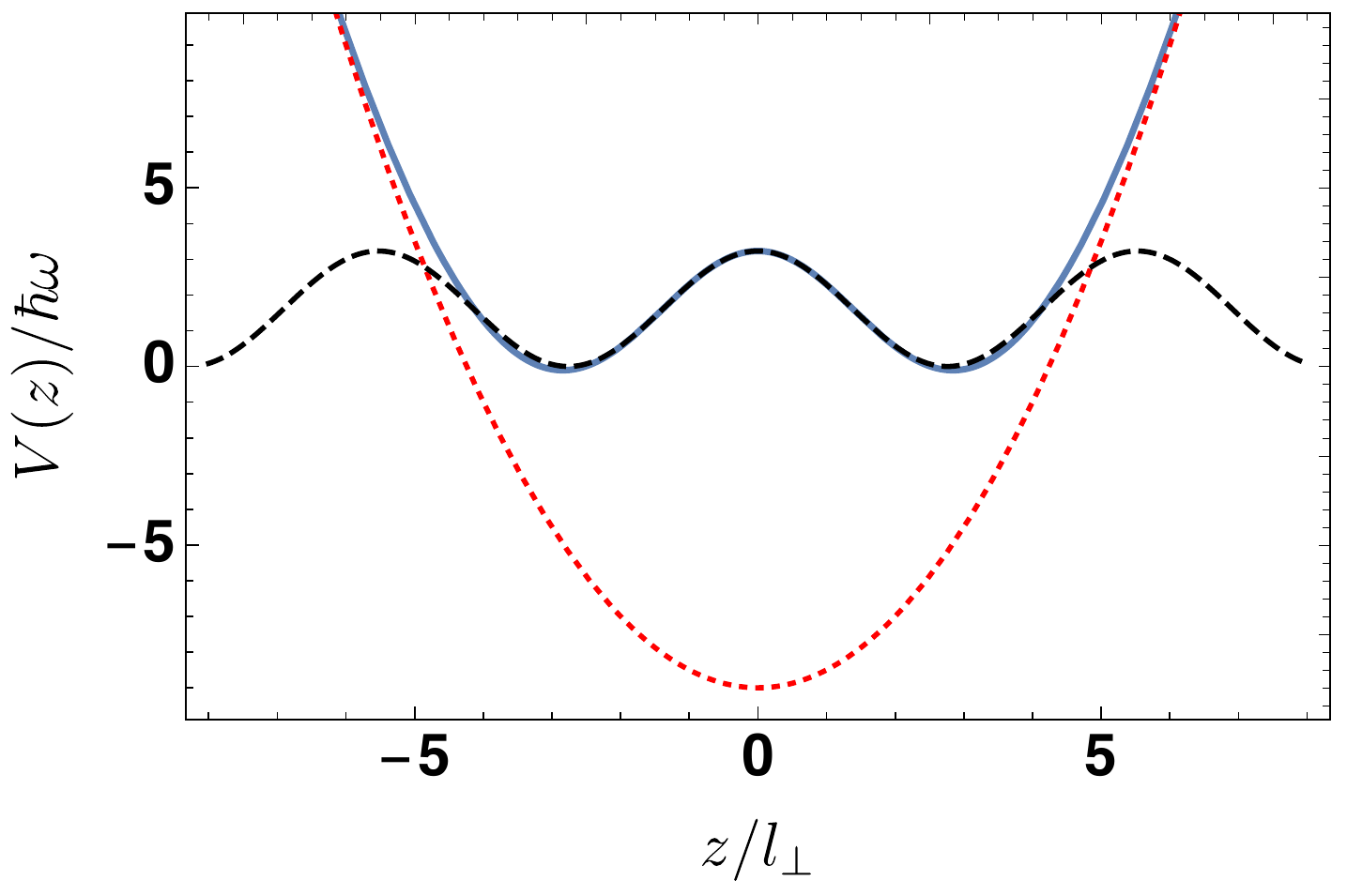}
	\caption{\label{setup}Trapping potential in the axial direction used here to mimick two sites of an optical lattice (straight blue line), compared with the actual lattice potential (dashed black line) and a purely harmonic trap (red dotted line).}
\end{figure}

The scheme described in the previous section can also be applied to the case in which the trapping potential is not fully harmonic. The only requirement is that the single particle energy spectrum becomes harmonic asymptotically at high energy~\cite{Jachymski2013a}. Here we are interested in the properties of two atoms trapped in two sites of an optical lattice characterized by the potential $V_L \cos^2 k_L x$, which are relevant for the experimental observations~\cite{Mark2018}. We thus use a simplified potential of the form
\beq
V(z)=V_L \left((z k_L)^2/a+\frac{4}{1+(z k_L)^2/b+(z k_L)^4/c}\right)
\eeq
with $\frac{1}{2}m\omega^2=V_L k_L^2/a$ setting the asymptotic harmonic trap frequency. In the radial direction  $V(\rho)=\frac{1}{2}m\eta^2\omega^2\rho^2$ with $\eta\gg 1$. The coefficients $a$, $b$ and $c$ are chosen such that the first few  terms of the series expansion of this potential agree with the expansion of the optical lattice potential. This allows to recover the double well structure. At the same time, the potential asymptotically reaches purely harmonic form at larger distance. This is illustrated in Figure~\ref{setup}. Note that here we focus on two identical atoms, which means that the harmonic trap frequency for the relative motion and the center of mass is the same. If needed, different trapping frequencies can be easily incorporated as we are anyway dealing with the center of mass to relative motion coupling induced by the second term of the potential.

With this choice of the potential we expect to be able to reproduce all essential processes taking place in a deep optical lattice, as the resonances between separated lattice sites are expected to be strongly suppressed due to very low tunnelling matrix element. For very precise studies it can be beneficial to make sure that more terms in the series expansion agree. In principle this computation method allows to study not only two or more lattice sites along a single direction, but also e.g. a $2\times2$ plaquette without much difficulty.

\section{Results and discussion}
\label{secres}
Having described the principles, we now turn to the discussion of an experimentally relevant example. We choose the atomic mass $m$ of $^{138}$Cs and the optical lattice depth $V_L=40E_R$ with the recoil energy $E_R=\hbar^2 k_L^2/(2m)$. This leads to the local harmonic oscillator frequency at a single site $\omega_L=16150\,$Hz and the asymptotic trap frequency $\omega$ of about $11170\,$Hz. The single particle trap eigenstates of a double well potential split into pairs of nearly degenerate states with even and odd symmetry. This is similar to the band structure of the full lattice potential, so we will refer to these pairs as bands. The trap anisotropy in the remaining two directions is set to $\eta=9.9$ so that transverse excitations do not matter.

\begin{figure}
	\includegraphics[width=0.45\textwidth]{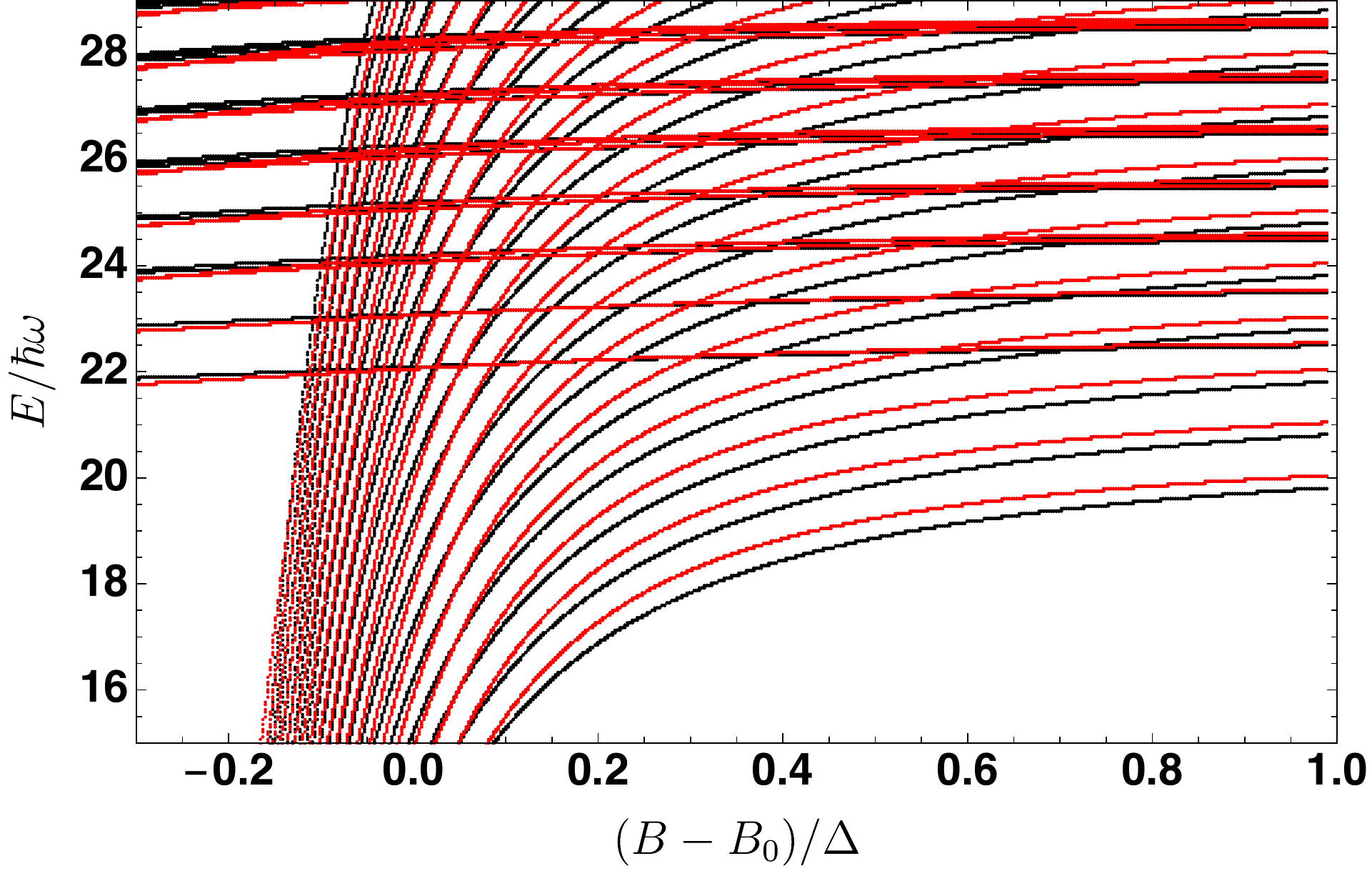}
	\caption{\label{fig:harm}Energy levels in a purely harmonic, anisotropic trap with frequency $\omega$ for the wide (black) and narrow (red or grey) Feshbach resonance as a function of magnetic field.}
\end{figure}

\begin{figure}
	\includegraphics[width=0.45\textwidth]{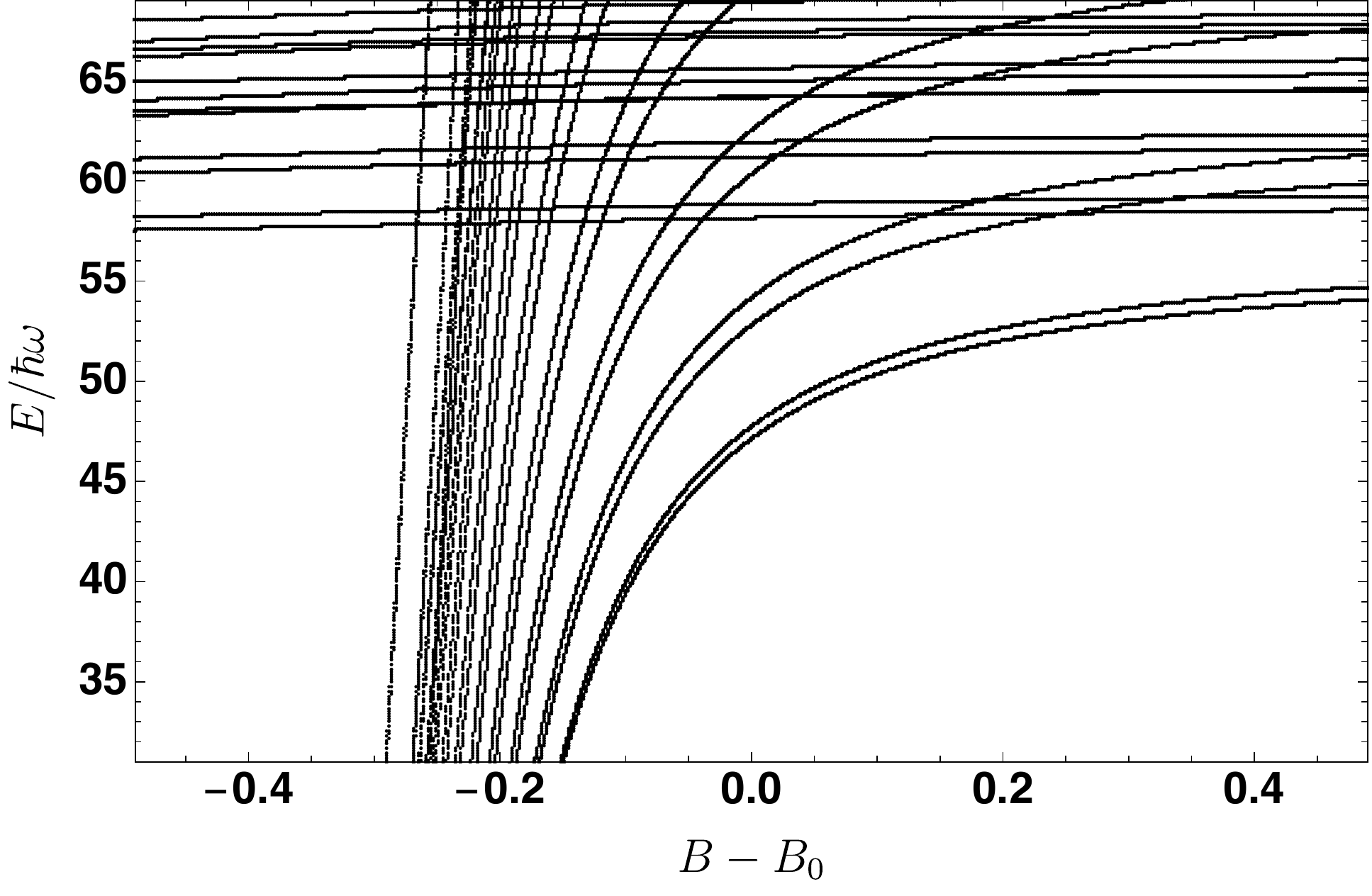}
	\caption{\label{fig:dw}Energy levels in a double well trap as a function of magnetic field for the wide Feshbach resonance. In contrast to the previous case, the molecular and trap states now in general exhibit anticrossings.}
\end{figure}

As an exemplary demonstration of the advantages of the double well trap we focus on studying the properties of two resonances. For the first one (later referred to as ``wide'') we set $\Delta=1$G , $a_{\rm bg}=120a_0$ and $\delta\mu=9\mu_B$ where $a_0$ is the Bohr radius and $\mu_B$ the Bohr magneton. For Cs this results in $s_6\approx 5.6$, which indicates an open-channel-dominated resonance. The second resonance (dubbed ``narrow'') has again $\Delta=1$G, but $a_{\rm bg}=90a_0$ , $\delta\mu=2\mu_B$, resulting in $s_6\approx 0.9$.  As shown in Figure~\ref{fig:harm}, in a harmonic trap these resonances give rise to extremely similar spectra which cannot be experimentally distinguished. 

\begin{figure}
	\includegraphics[width=0.45\textwidth]{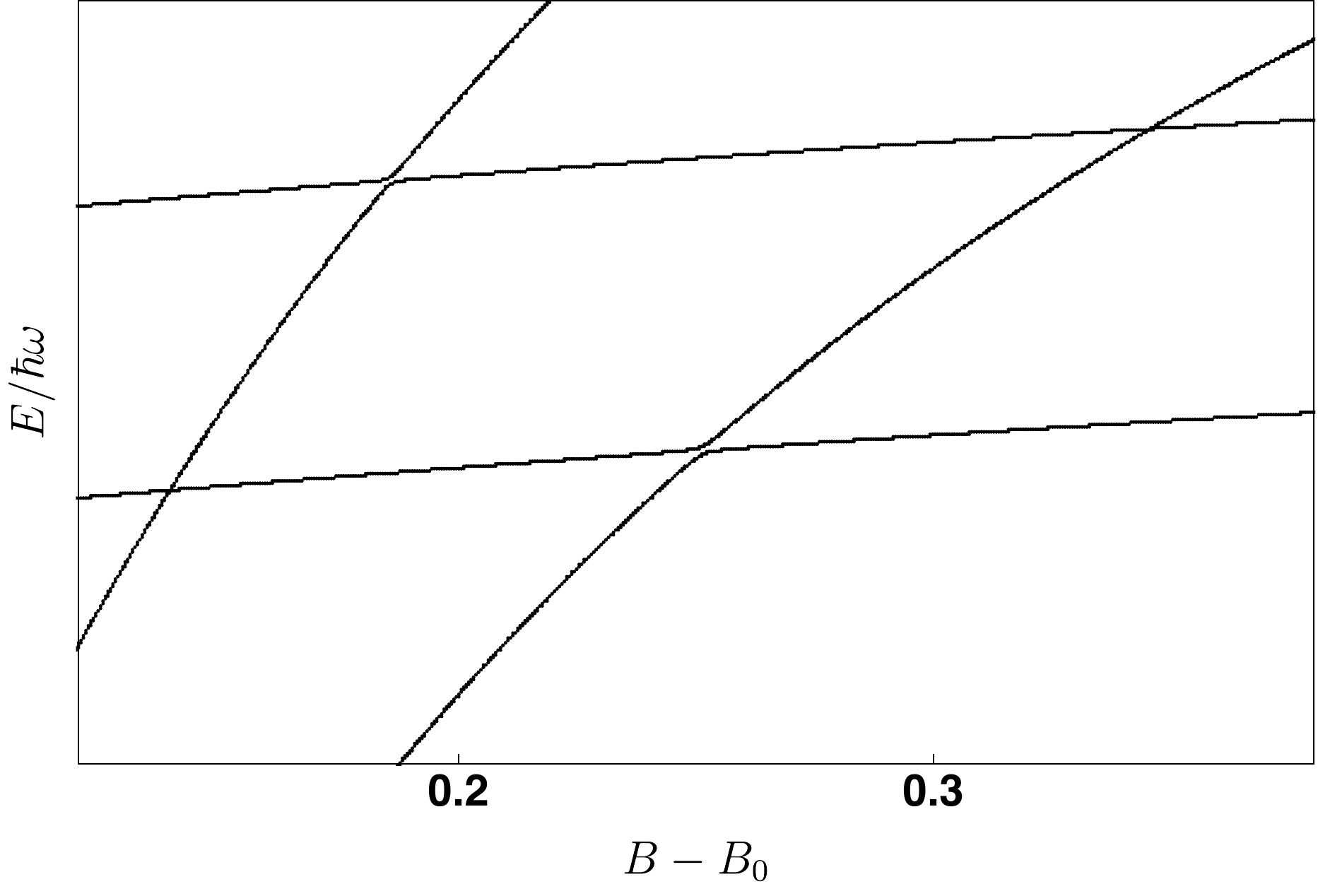}
	\includegraphics[width=0.45\textwidth]{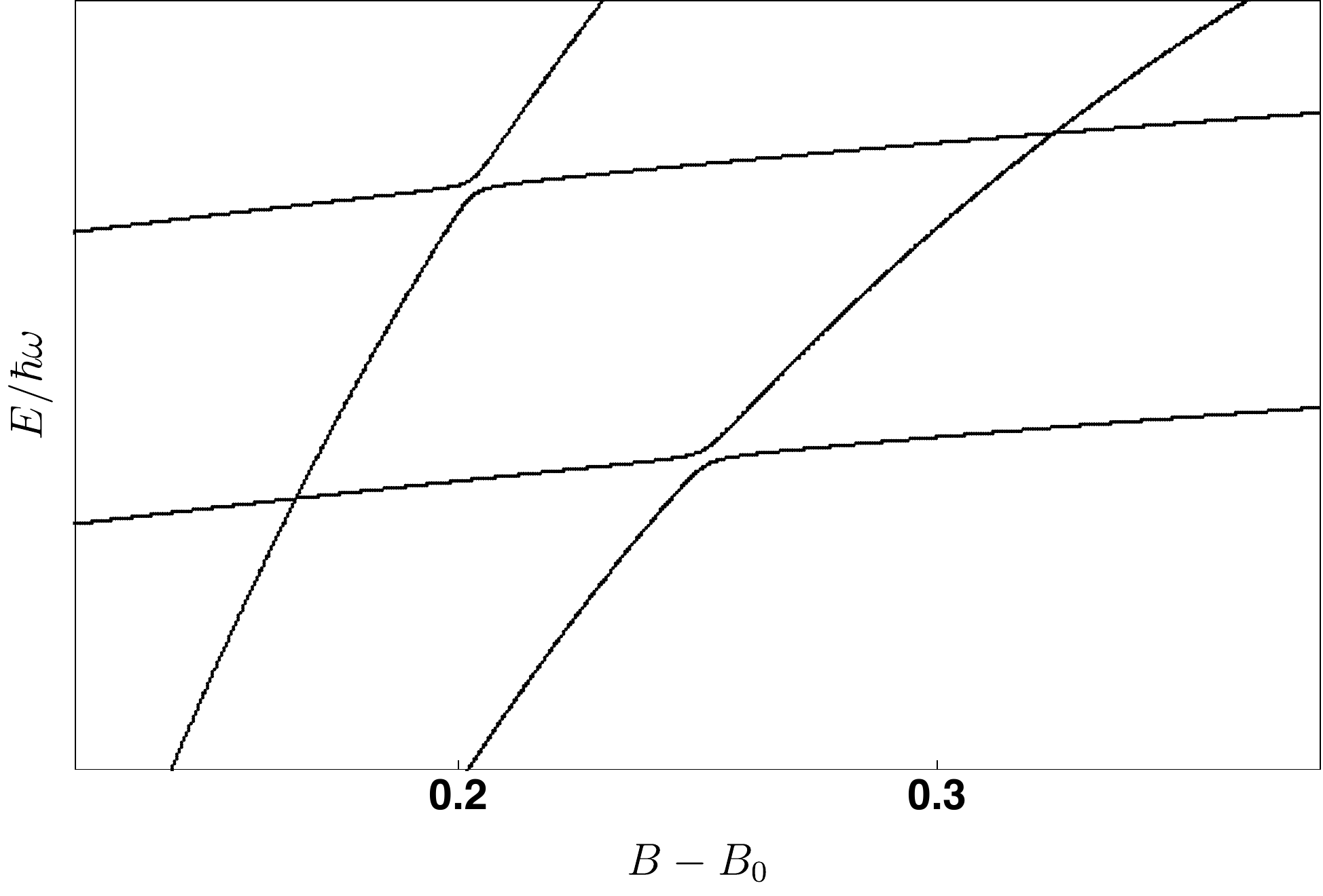}
	\caption{\label{fig:cross}Upper: part of the energy spectrum shown in Fig.~\ref{fig:dw} focusing on the intersection between the first excited molecular states and the lowest lattice band. Lower: the same crossing, but for the case of the narrow resonance.}
\end{figure}

\begin{figure}
	\includegraphics[width=0.45\textwidth]{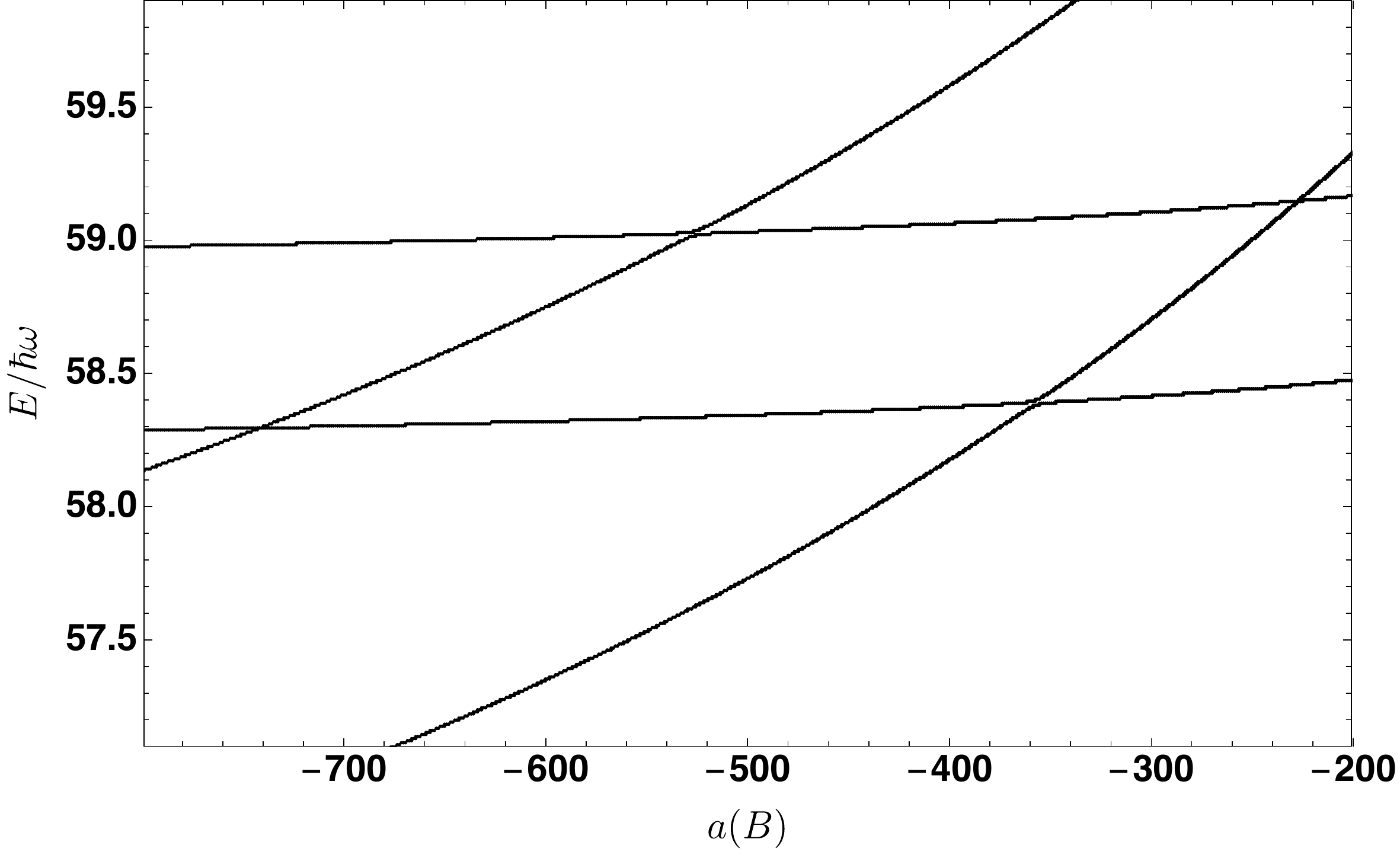}
	\includegraphics[width=0.45\textwidth]{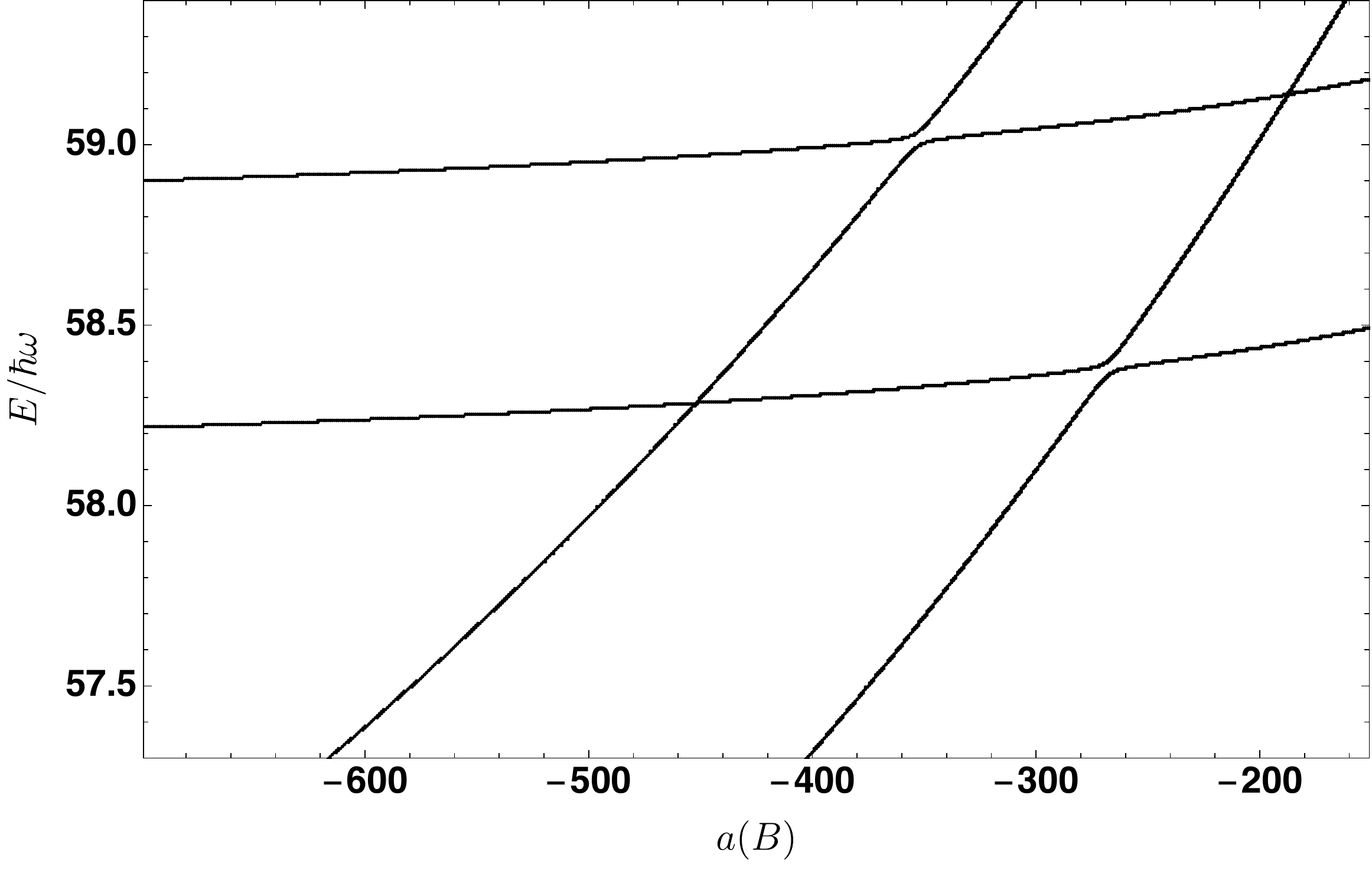}
	\caption{\label{fig:cscat}The same crossing as in Fig.~\ref{fig:cross} plotted as a function of the scattering length.}
\end{figure}

Let us now turn to the case of the double well trap displayed in Figure~\ref{fig:dw}. Here, similarily to the previous case, the ramping states are the molecular states which intersect with slowly varying trap states. Each molecular level approaches the respective trap state with the same symmetry and number of center of mass excitations. However, due to the coupling of center of mass and relative motion, the excited molecular states have nonzero overlap with the trap-bound states. This is shown in detail in Figure~\ref{fig:cross} which highlights the crossing of the first excited molecular levels with the lowest band. Each level exhibits a crossing with the state of different symmetry and an anticrossing if the symmetry matches. These anticrossings can cause population transfer to the molecular state as the magnetic field is varied, which leads to additional loss features in experimental measurements. Quite strikingly, the positions of the crossings depend strongly on the resonance parameters, as can be seen by comparing the two resonances shown in Fig.~\ref{fig:cross}. One can also plot the energy levels as a function of the scattering length which is the more convenient way of comparing resonances with different magnetic field width $\Delta$, as done in Fig.~\ref{fig:cscat}. For the chosen parameters, in the case of the wide resonance one can expect additional features at about $-360$G and $-530$G, while for the narrow resonance the anticrossing occur at $-270$ and $-380$G. The shifts becomes more significant as the difference in character of the resonances described by $s_6$ is increased.

A simple model can qualitatively explain the observed nonuniversal (in the sense that they depend not only on the scattering length) shifts. Consider a single molecular state getting shifted only due to coupling to a single level. We are looking for the impact of the coupling on the position of the crossing with the first band. In the first order, the shift is then proportional to $g^2$, which is directly proportional to $s_6$ (see eq.~\eqref{gdef}), in agreement with the results of Fig.~\ref{fig:cscat}. One can also consider the shift as an effect of the effective range of the interaction potential, which for our model is given by~\cite{Werner2012} $r_{\rm eff}=-\frac{\hbar^2}{\mu a_{\rm bg}\Delta \delta\mu}(1-\frac{a_{\rm bg}}{a(B)})^2$, being negative and inversely proportional to $s_6$. This clearly shows the necessity of using the two-channel model. In the case of a single channel treatment using a model interaction such as the Lennard-Jones potential, one obtains for the effective range a universal formula~\cite{Gao1998} $r_{vdW}=\bar{a}\frac{\Gamma(1/4)^2}{6\pi^2}\left(1-\frac{2\bar{a}}{a(B)}+\frac{2\bar{a}^2}{a(B)^2}\right)$. This small correction is not present in our model due to the lack of explicit background interaction term, but is negligible close to the resonance and can be incorporated if needed.

\begin{figure}
	\includegraphics[width=0.5\textwidth]{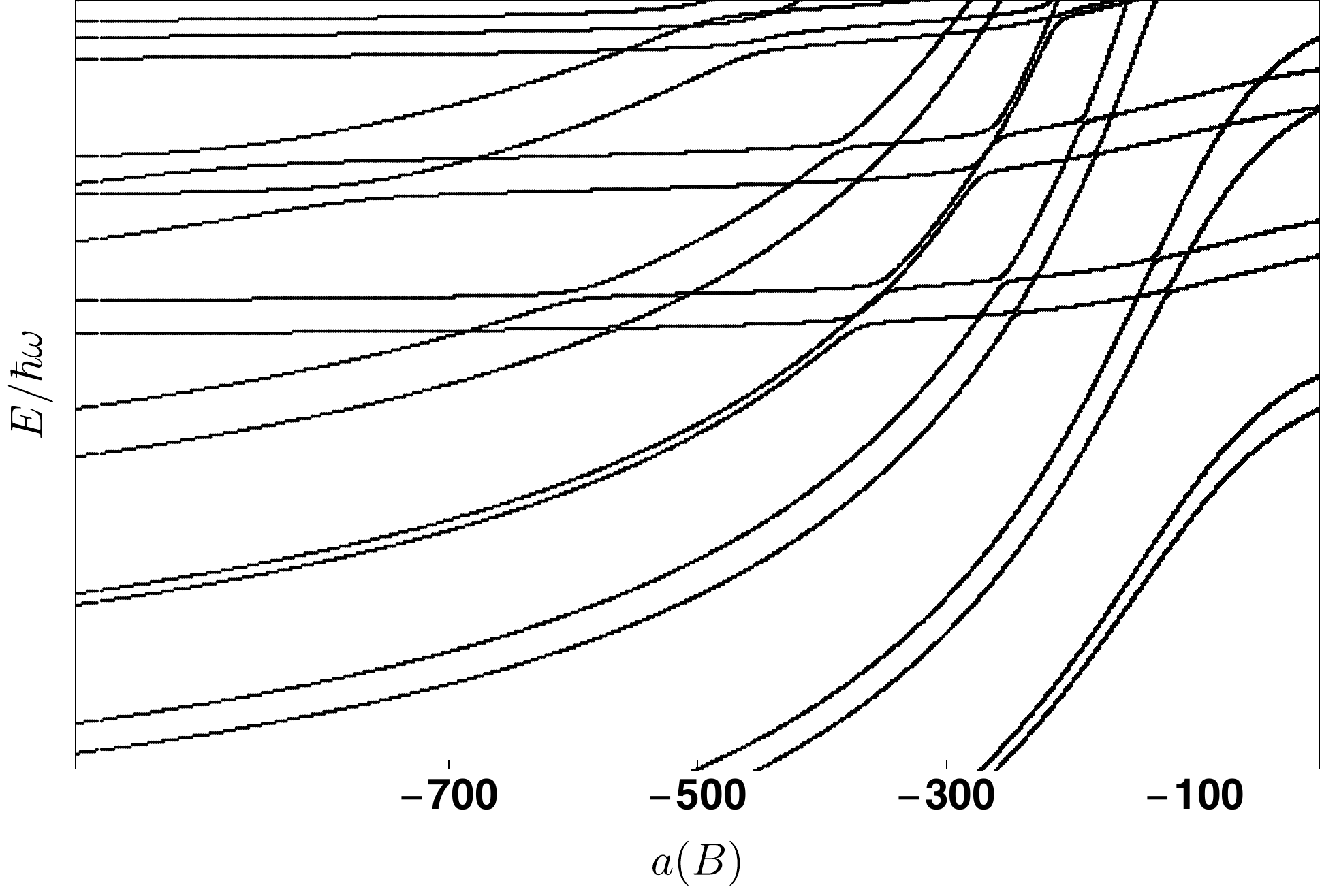}
	\caption{\label{fig:narrow}Crossings of the molecular levels with the first two lattice bands for a closed-channel-dominated resonance.}
\end{figure}

In the experimental study of Feshbach resonances reported by~\cite{Mark2018} the atoms are initially prepared in a Mott insulator state with singly occupied sites and the number of atoms is measured after holding the sample at certain magnetic field. Losses are only possible if the atoms are allowed to tunnel within the lattice to undergo recombination. For the perfect scenario this occurs when the interaction is switched off at the zero crossing of the scattering length. In the experiment, gravity introduces a tilt in the onsite energies, shifting the position of the insulator melting point. As described here, additional loss features are caused by the coupling to excited molecular states with their positions depending on the resonance pole strength. This process leads to association of a molecule from two atoms placed initially in separate lattice sites.

Finally, let us briefly consider the case of a strongly closed-channel-dominated resonance. Such situation implies that the trap states are heavily admixtured with the molecular component near a crossing. This leads to strong repulsion between the levels, modifying the curvature of the ramping states. This is illustrated in Fig.~\ref{fig:narrow} for the case of  $a_{\rm bg}=20a_0$ and $\delta\mu=1.5\mu_B$, which gives $s_6\approx 0.15$. Here visible anticrossings occur even for the higher lattice band and excited molecular states, which would lead to observation of a complicated loss spectrum.

\section{Conclusions}
\label{secconc}

We have demonstrated the benefits of using tight external traps for studying Feshbach resonances. Apart from providing much cleaner spectra than free space spectroscopy by eliminating the three-body losses, the trap enables more complete characterization than just the resonance position and width. By measuring the additional resonances induced by the coupling of the center of mass and relative motion, it is possible to estimate the pole strength of the resonance and track down the energy dependence of the scattering phase shift. This can be relevant not only for more precise spectroscopy and testing different atomic structure models as well as studying the validity of common approximations, but also for quantum simulations of many-body  physics with cold atoms, as the molecular fraction can be crucial for the system properties. 

In this work, we focused on the properties of a double well system which can be relevant both for optical lattices and for tweezer systems. In an actual optical lattice one would expect multiple more resonances coming from the three-dimensional structure as well as from further lattice sites. The calculations can be extended to include these features, but we expect additional resonances in an anisotropic lattice to be separated from the ones studied here, and to be far more narrow (such as the crossing between the lowest band and the third molecular state visible in Fig.~\ref{fig:dw}). A probably more significant extension to the current work would be to include the background interaction in the open channel, which would lead to an additional bound state that can be relevant if the background scattering length is large.

The author would like to thank Florian Meinert and Shmuel Fishman for inspiring discussions.

\bibliography{Allrefs}

\begin{thebibliography}{44}
\bibitem [{vdW()}]{vdW}%
  \BibitemOpen
  \href@noop {} {}\bibinfo {note} {The van der Waals length and energy can be
  defined as $R_6=(2\mu C_6/\hbar^2)^{1/4}$ and $E_6 = \hbar^2/(2\mu R_6^2)$
  with $C_6$ being the van der Waals interaction coefficient; typically, $R_6
  \sim 100 a_0$.}\BibitemShut {Stop}%
\end{thebibliography}
\end{document}